\documentclass[3p,times]{elsarticle}

\usepackage{ecrc}
\usepackage{lineno}
\volume{00}

\firstpage{1}

\journalname{Nuclear Physics A}

\runauth{}

\jid{nupha}

\usepackage{amssymb}

\usepackage[figuresright]{rotating}

\begin{document}

\begin{frontmatter}



\dochead{}

\title{Open charm hadron production via hadronic decays at STAR}


\author{\small David Tlust\'y for the STAR collaboration}

\address[ujf]{Nuclear Physics Institute, Academy of Sciences Czech Republic, Na Truhl\'a\v{r}ce 39/64, 180 86 Praha 8, Czech Republic}
\address[cvut]{Czech Technical University in Prague, Faculty of Nuclear Sciences and Physical Engineering, B\v{r}ehov\'a 7, 11519, Prague 1, Czech Republic}

\begin{abstract}
Heavy quarks are a unique probe to study the medium produced in ultra-relativistic heavy ion collisions. The dominant process of charm quark production at RHIC is believed to be initial gluon fusion which can be calculated in the perturbative QCD. The upper limit of FONLL calculation seems to be in good agreement with charm cross section measurements at mid-rapidity in $p+p$ collisions at $\sqrt{s_{NN}}$ = 200 GeV provided by STAR. The same measurement in Au+Au collisions at equal energy reveals the number-of-binary-collisions scaling of charm cross section indicating that charm production is dominated by initial hard scatterings. In this article, we report the measurements of $D^{0}$, $D^{*}$ in $p+p$ at 0.6 GeV/$c < p_T < 6$ GeV/$c$ and $D^0$ in Au+Au collisions at 0.2 GeV/$c < p_T < 5$ GeV/$c$ via hadronic decays $D^{0}\rightarrow K^-\pi^+,\ D^{*+}\rightarrow D^0\pi^+\rightarrow K^-\pi^+\pi^+$  at mid-rapidity $|y|<1$. \end{abstract}


\end{frontmatter}


\section{Introduction}
\label{sec:intro}

The charm quark production is dominated by initial gluon fusion at initial hard partonic collisions 
and well described by perturbative QCD (pQCD) because of large charm quark mass ($\sim$ 1.5 GeV/c$^2$) \cite{cprod}. The mass is almost exclusively generated through its coupling to the Higgs field in
the electroweak sector, while masses of (u, d, s) quarks are dominated by spontaneous breaking of
chiral symmetry (CS) in QCD \cite{CharmMass}. This means that charm quarks remain heavy even if CS
is restored, as it likely is in a QGP. Since charm is produced mainly in initial hard scatterings, one expects
that charm production total cross section
$\sigma^{NN}_{c\bar{c}}$ should scale as a function of number-of-binary-collisions $N_{\mathrm{bin}}$. 
In addition, if charm quarks flow elliptically, there must have been enough interactions to easily thermalize light quarks. Hence, charm quarks are an ideal probe to study early dynamics in high-energy nuclear collisions.      
 
\section{Measurement}
\label{sec:measurement}

\begin{figure}[!h]
\begin{center}
\vspace{-1mm}
\includegraphics[width=0.7\textwidth]{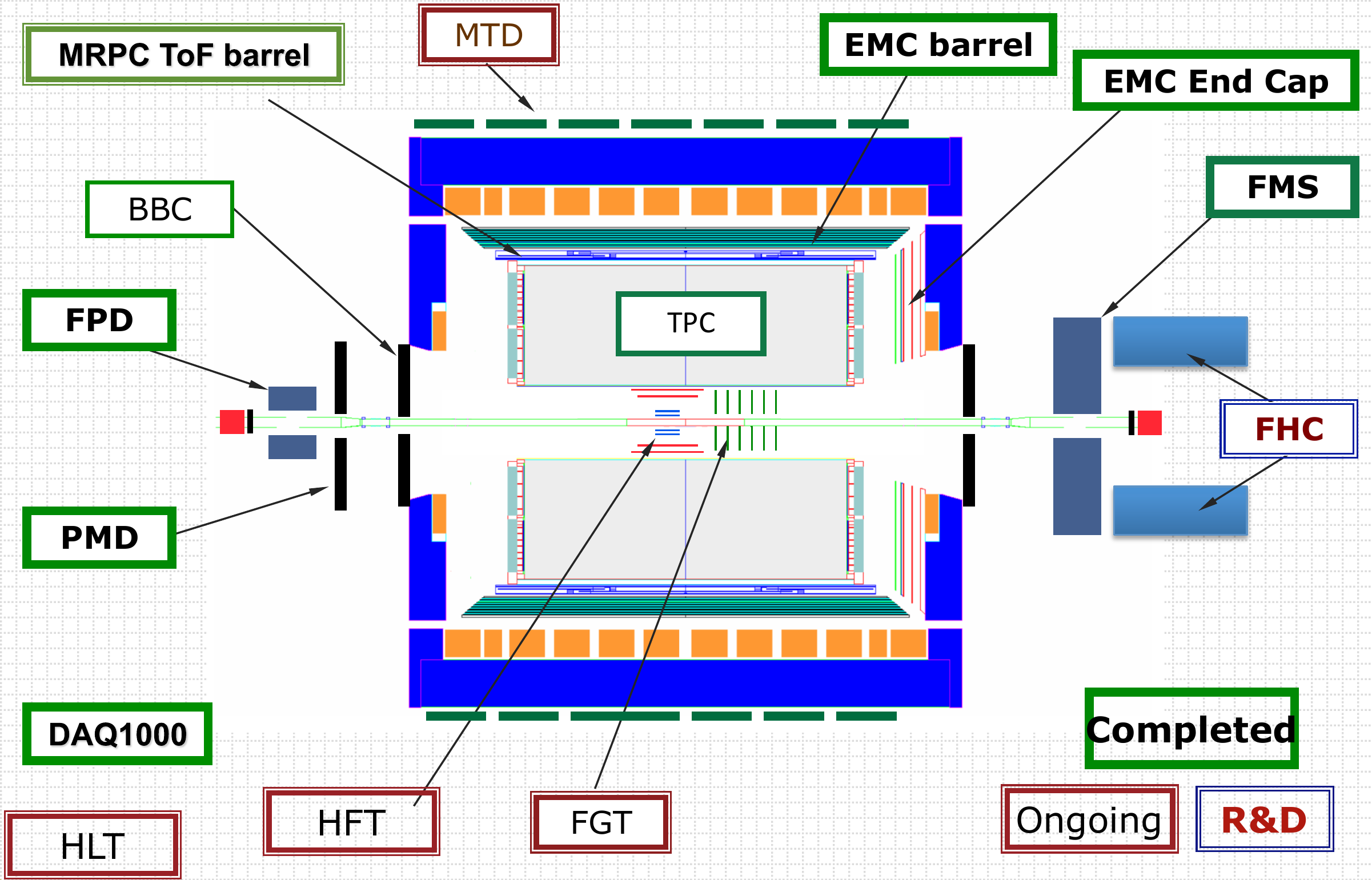} 
\end{center}
\vspace{-3mm}
\caption{The STAR detector. TPC (Time Projection Chamber) is main detector
for tracking and PID (providing $\mathrm{d}E/\mathrm{d}x,\vec{p})$, TOF (Time Of Flight) for PID improvement,
and pile-up tracks removal and BEMC (Barrel Electromagnetic
Calorimeter) in this analysis for pile-up tracks removal if TOF is not available.} 
\label{star}
\end{figure}

Charm cross section at mid-rapidity is calculated from open charm hadrons yields. These yields  
are obtained from the invariant mass reconstruction of open charm mesons 
through hadronic decays: $D^0(\overline{D^0})\rightarrow K^\mp\pi^\pm$ (BR = 3.89\%) and $D^{*\pm}\rightarrow D^0(\overline{D^0})\pi^\pm$ (BR = 67.7\%) $
\rightarrow K^-\pi^+\pi^\pm$. The identification of daughter particles was done in the STAR experiment (see Fig. \ref{star}) at
mid-rapidity $|y|<1$ at $\sqrt{s_{NN}}=200$ GeV.

The analysis presented herein was done using two datasets; the first one collected in 2009 with minimum bias trigger defined as a coincidence in Vertex Position Detectors East and West (105 million events total) and the second one collected in 2010 with minimum bias trigger defined as a signal in Zero Degree Calorimeter (280 million events total).

Sub-detectors used in this analysis were: 
\begin{itemize}
\item Time Projection Chamber (TPC) providing 3D image of a particle track and ionization energy loss $(\mathrm{d}E/\mathrm{d}x)$ and covering a large acceptance with full $2\pi$ azimuthal angle at $|\eta| < 1$.
With the help of uniform magnetic field one can calculate momentum vector of a track as a function of
track helix radius and magnetic induction.  
\item Time-Of-Flight (TOF) detector providing time of the particle's flight from primary vertex to a channel
pad and covering 72\% in 2009 and 100\% in 2010 of the whole barrel. When combined with momentum measurements
from the TPC, this allows to separate pions from kaons up to 1.6 GeV/c \cite{NIM}. 
\end{itemize}

STAR doesn't have any subsystem being able to reconstruct secondary vertex of $D^0$ or $D^*$ decay; one must
calculate invariant mass of all $K\pi$ pair coming from vicinity of primary vertex. This results in a large 
combinatorial background which was reconstructed via mixed-event method (Au+Au dataset), same-charge-sign, and kaon momentum-rotation (p+p dataset) and subtracted from invariant mass spectra of daughter particle pairs \cite{OpenCharmPrev}. To reconstruct
$D^*$, one may exploit the softness of $D^*\rightarrow D^0\pi$ decay; combine low momentum pions with $D^0$
candidates, i.e. pairs with $1.82<M(K\pi)<1.9$ GeV/c, and plot difference $M(K\pi\pi)-M(K\pi)$.
Such value occupies the beginning of the phase space; hence it does not suffer from
large combinatorial background making $D^*$ direct observation possible. The combinatorial background was
reconstructed by side-band (picking $D^0$ candidates outside the $D^0$ mass region) and wrong-sign (picking
soft pion with opposite charge) methods. The difference between these methods is the dominant source of systematic uncertainties for both $D^0$ and $D^*$ analyses.         

\begin{figure}[!h]
\begin{center}
\vspace{-1mm}
\includegraphics[width=0.30\textwidth]{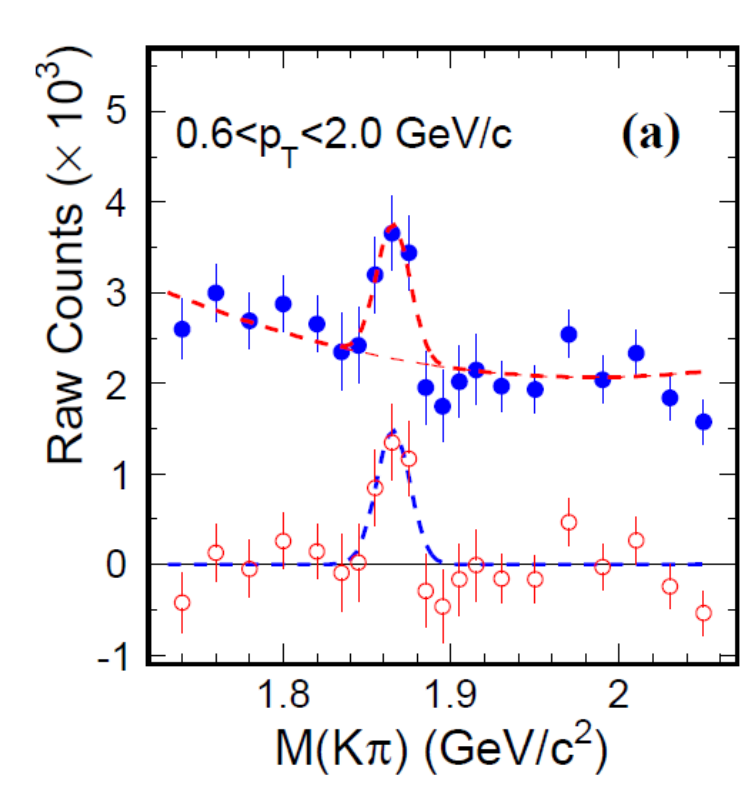} 
\includegraphics[width=0.30\textwidth]{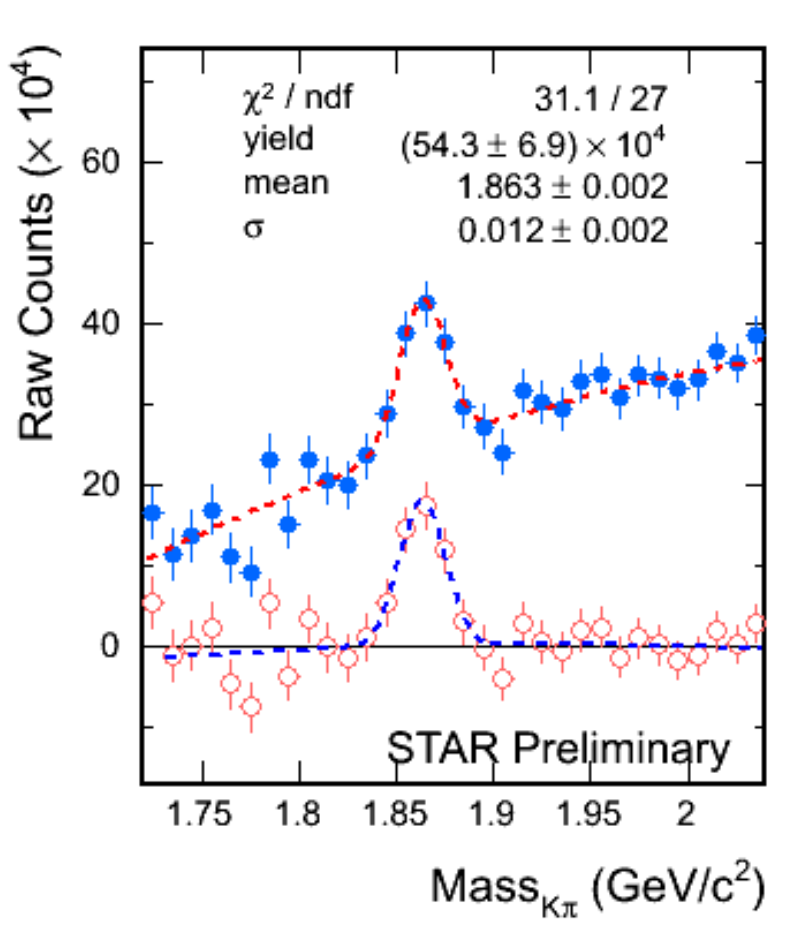}
\includegraphics[width=0.39\textwidth]{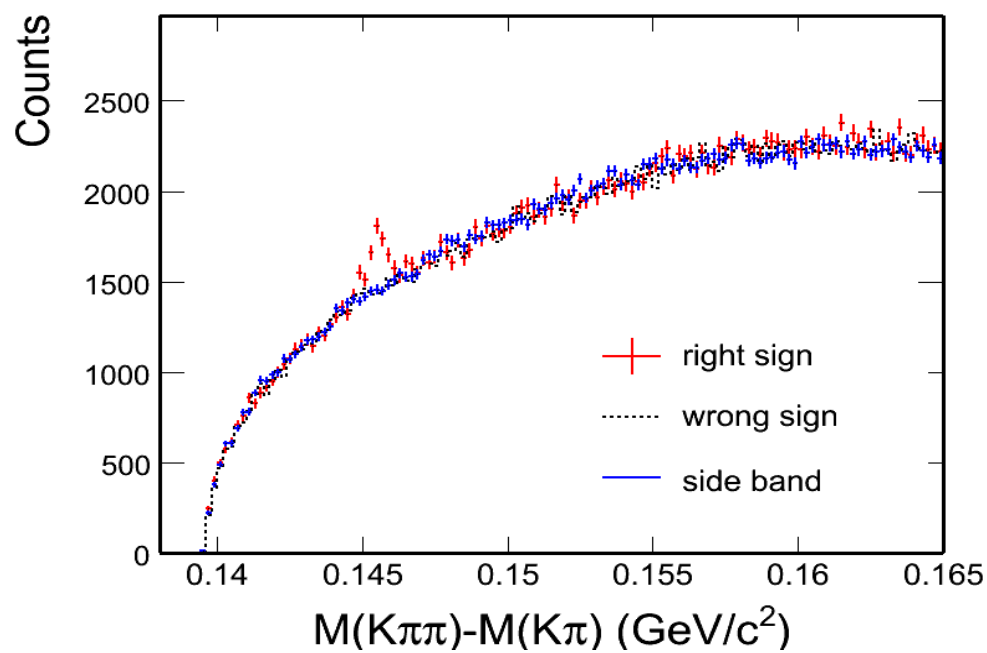}
\end{center}
\vspace{-3mm}
\caption{Left panel: $D^0$ signal in p+p 200 GeV  
collisions after same-sign background subtraction. Middle panel: $D^0$ signal in Au+Au 200 GeV  
collisions after mixed-event background subtraction Right panel: $D^*$ 
signal in p+p 200 GeV collisions. Combinatorial background is reproduced by the 
distributions from the wrong-sign (black dotted) and side-band (blue solid) methods. } 
\label{yields}
\end{figure}

Open charm hadrons yields $Y_{D^0,D^*}$ were calculated as Gaussian 
function areas from the invariant mass peaks fits shown in Figure \ref{yields}. Raw counts 
were corrected with the reconstruction efficiency in used sub-detectors to obtain the correct $Y_{D^0,D^*}$.

\section{RESULTS}

\begin{figure}[!h]
\begin{center}
\vspace{-1mm}
\includegraphics[width=0.56\textwidth]{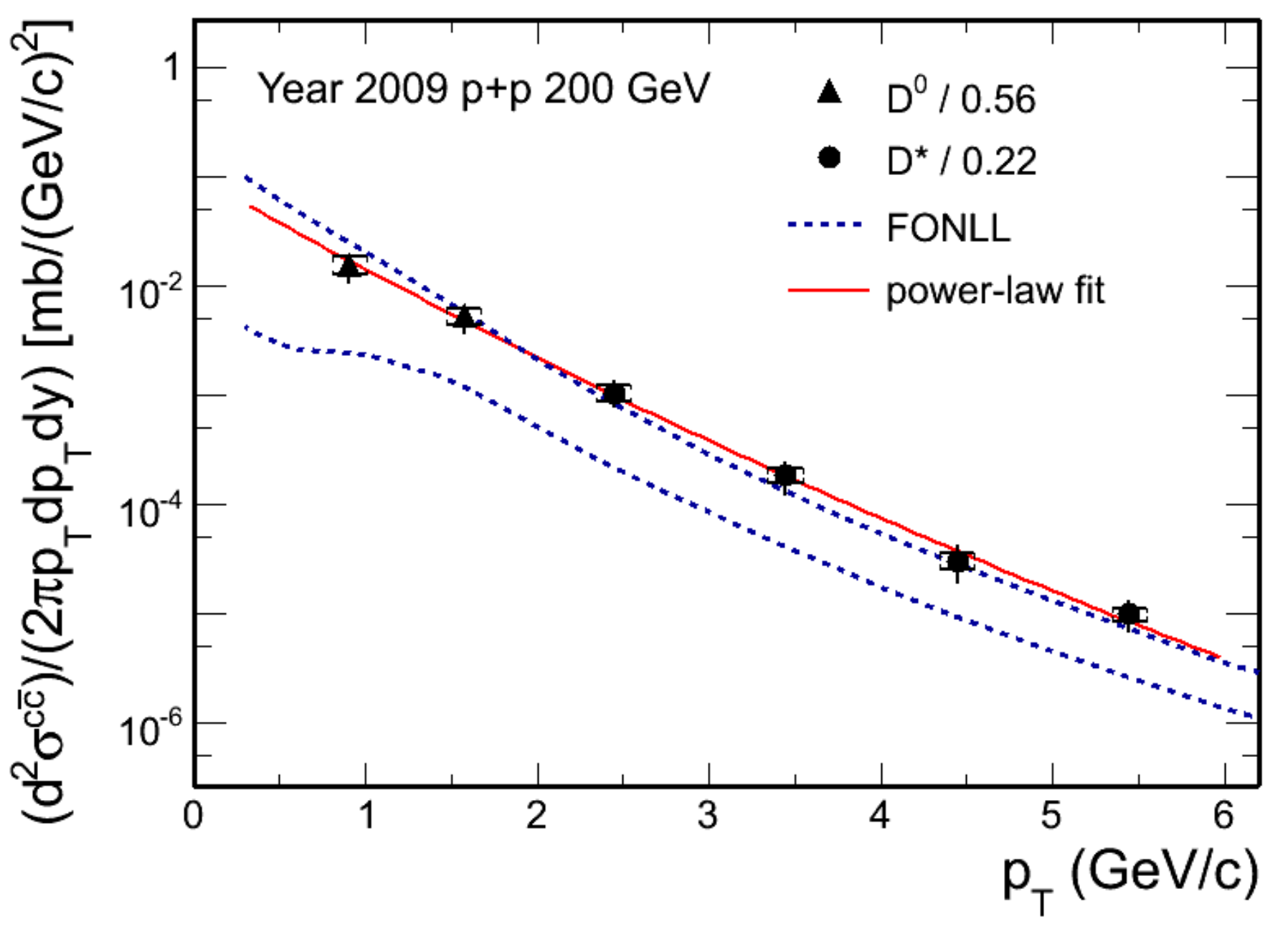} 
\includegraphics[width=0.43\textwidth]{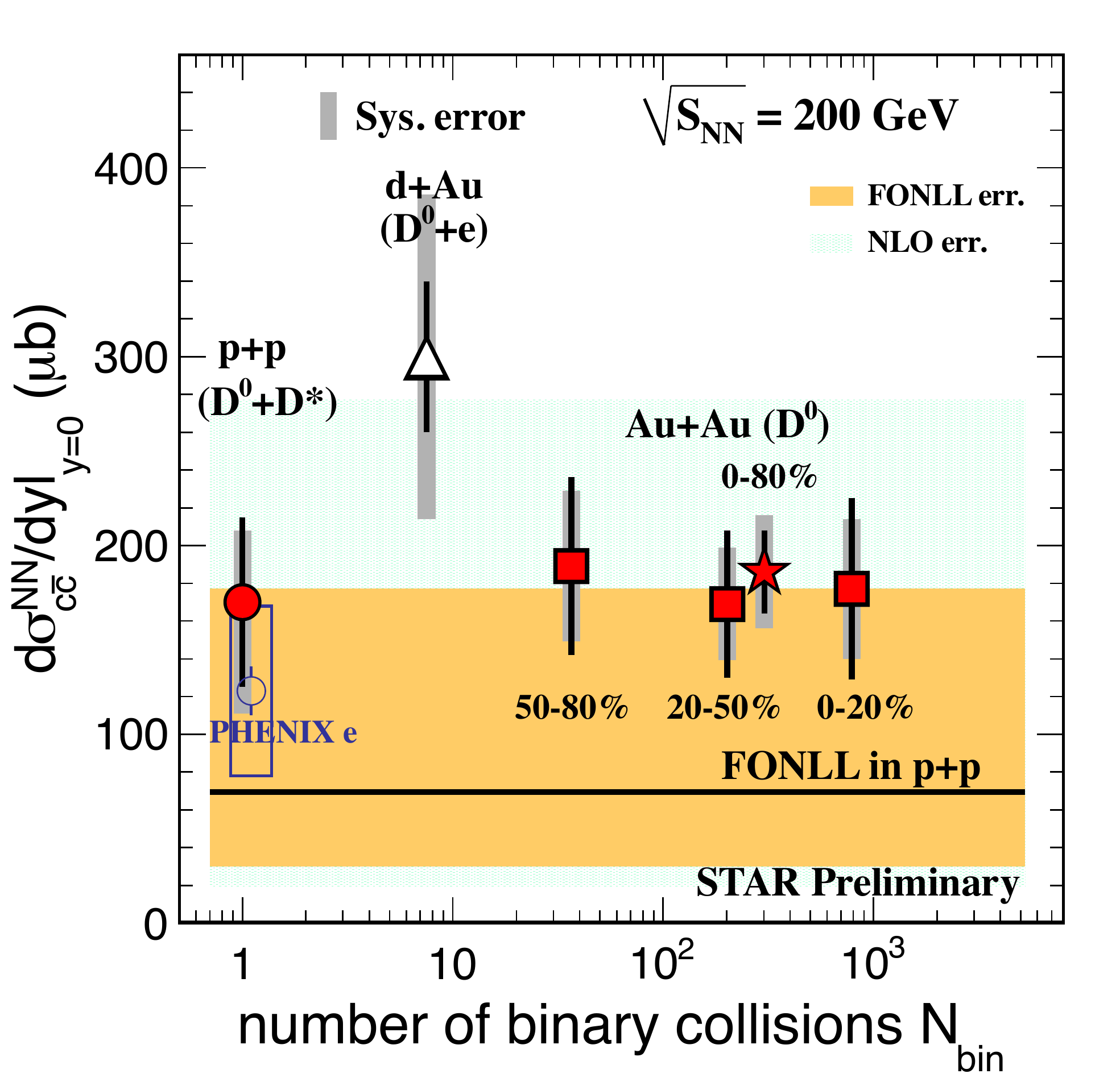} 
\end{center}
\vspace{-3mm}
\caption{Left Panel: $c\overline{c}$ pair production cross section (symbols) as a function of 
$p_T$ in 200 GeV p+p collisions. Right Panel: The charm production cross section per 
nucleon-nucleon collision at mid-rapidity as a function of $N_{\mathrm{bin}}$.} 
\label{spectrum1}
\end{figure}

Yields were calculated in 6 $p_T$ bins (2 for $D^0$, 4 for $D^*$) in p+p dataset and 5 $p_T$ bins (all for $D^0$) in 
Au+Au dataset.
Then the invariant charm cross section $\mathrm{d}\sigma_{\mathrm{p+p}}^{c\overline{c}}/(2\pi p_T\mathrm{d}p_T\mathrm{d}y)$ was
calculated by formula
\begin{equation}
\frac{\mathrm{d}\sigma^{c\overline{c}}}{2\pi p_T\mathrm{d}p_T\mathrm{d}y}=\frac{Y\sigma^{\mathrm{NSD}}}{2\pi p_T\Delta p_T \Delta y\, \mathrm{BR} \, }\frac{\epsilon_\mathrm{T}}{N f_\mathrm{frag.}} 
\label{xsection} 
\end{equation}
in each $p_T$ bin ($\sigma^{\mathrm{NSD}}$ is non-single diffractive p+p inelastic cross section, $f_\mathrm{frag.}$ is the ratio of charm quarks hadronized to open charm mesons and $\epsilon_\mathrm{T}$ is
the trigger bias correction). The charm cross section at mid rapidity $\mathrm{d}\sigma^{c\overline{c}}/\mathrm{d}y$ was
obtained from power-law function fit to $\mathrm{d}\sigma^{c\overline{c}}/(2\pi p_T\mathrm{d}p_T\mathrm{d}y)$ points (see Fig. \ref{spectrum1}) as $170\pm45(\mathrm{stat.})^{+37}_{-51}(\mathrm{sys.})\ \mu\mathrm{b}$. In Au+Au collisions we
calculated invariant yield $\mathrm{d}^2N/(N_{\mathrm{ev}}p_T\mathrm{d}p_T\mathrm{d}y)$.

The $\mathrm{d}\sigma^{c\overline{c}}/\mathrm{d}y$ at mid-rapidity in Au+Au collisions 
was extracted, from a power-law fit as $186\pm22(\mathrm{stat.})\pm30(\mathrm{sys.})\pm18(\mathrm{norm.})\ \mu\mathrm{b}$ 
assuming that the $f_{\mathrm{frag.}}$ does not change from p+p to Au+Au collisions. The charm cross section for three centrality bins, 0-20\%, 20-50\% 
and 50-80\%, is obtained according to the integrated yields. The charm production cross 
section per nucleon-nucleon collision at mid-rapidity as a function of $N_{\mathrm{bin}}$ is shown in 
the right panel of Fig. \ref{spectrum1}. Within errors, the results are in agreement and follow the 
number of binary collisions scaling, which indicates that charm quark is produced via 
initial hard scatterings at early stage of the collisions at RHIC. The FONLL (orange 
band) and NLO \cite{nlo} (light-blue band) uncertainties are also shown here for comparison.

\begin{figure}[!h]
\begin{center}
\vspace{-1mm}
\includegraphics[width=0.4\textwidth]{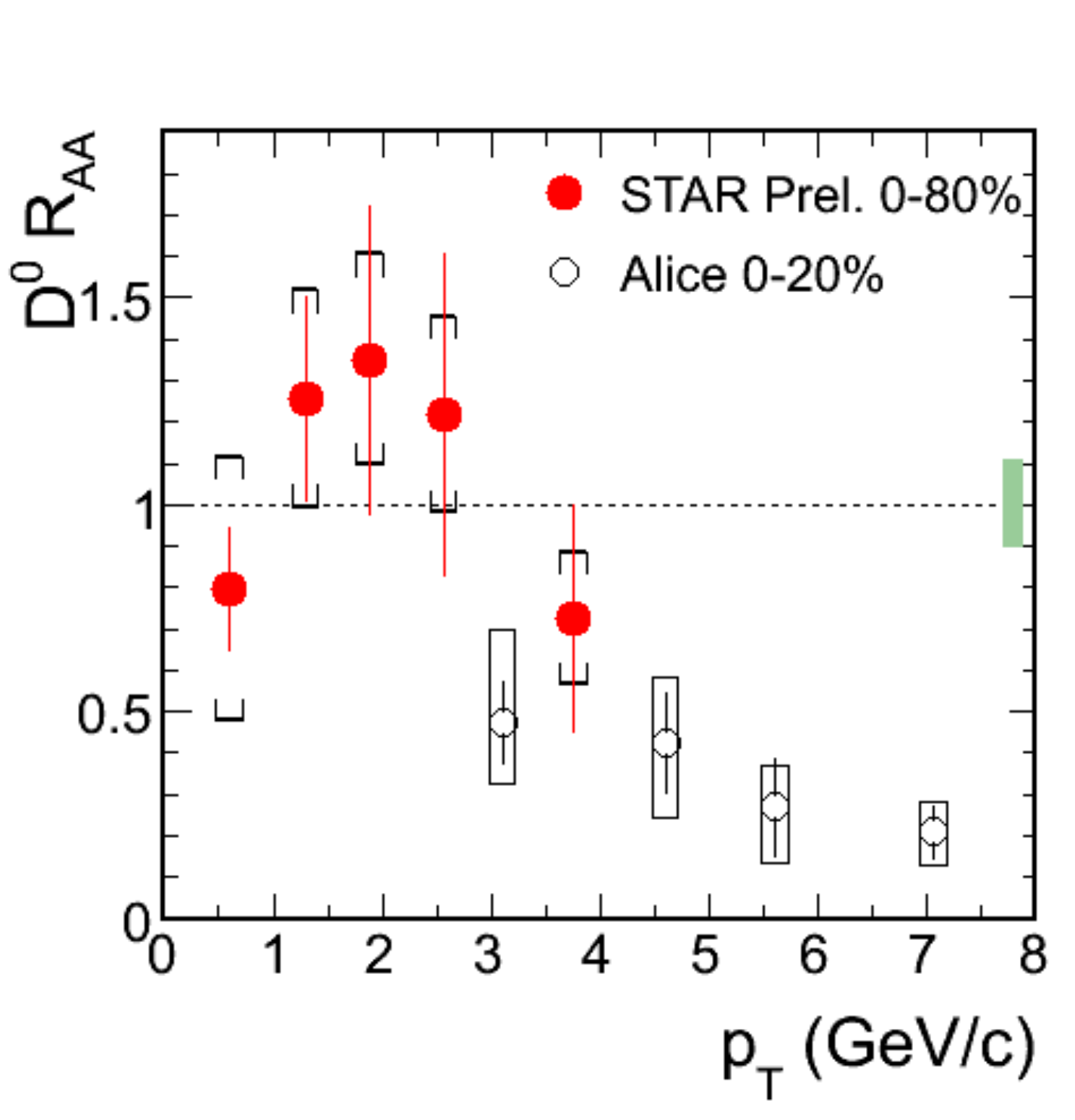} 
\end{center}
\vspace{-3mm}
\caption{$D^0$ nuclear modification factor $R_{\mathrm{AA}}$ as a function of p. ALICE data taken from \cite{AliceRAA}} 
\label{spectrum2}
\end{figure}

The $R^{D^0}_{\mathrm{AA}}$ depicted in Figure \ref{spectrum2} was obtained via dividing $D^0$ yields in 
Au+Au 0-80\% minbias collisions by the power-law fit to p+p yields scaled by $N_{\mathrm{bin}}$, 
The uncertainty of the p+p power-law shape is taken into account as systematic error. No suppression is observed at $p_T<3$ GeV/c.

\section{CONCLUSIONS}

Open charm hadrons $(D^0, D^{*+})$ are measured in p+p and Au+Au collisions at $\sqrt{s_{\mathrm{NN}}}=200$ GeV at STAR. Charm cross sections per nucleon-nucleon collision at mid-rapidity follow the number of binary collisions scaling. The charm pair production cross sections per nucleon-nucleon collision at mid rapidity are measured to be
$\mathrm{d}\sigma^{c\overline{c}}/(2\pi p_T\mathrm{d}p_T\mathrm{d}y)=170\pm45(\mathrm{stat.})^{+37}_{-51}(\mathrm{sys.})\ \mu\mathrm{b}$ in p+p and
$186\pm22(\mathrm{stat.})\pm30(\mathrm{sys.})\pm18(\mathrm{norm.})\ \mu\mathrm{b}$ in Au+Au minimum bias collisions. 
In the near future the STAR Heavy Flavor Tracker \cite{hft} will provide the 
necessary resolution to reconstruct secondary vertices of charm mesons, which will increase 
the precision of charm measurements. 

\textbf{Acknowledments}

This work was supported by grant INGO LA09013 of the Ministry of Education, Youth and Sports
of the Czech Republic, and by the Grant Agency of the Czech Technical University in Prague, grant No.
SGS10/292/OHK4/3T/14.




\bibliographystyle{elsarticle-num}
\bibliography{<your-bib-database>}



\end{document}